# A single chip 1.024 Tb/s silicon photonics PAM4 receiver

Ali Pirmoradi[†], Han Hao[†], Kaisarbek Omirzakhov, Alexander J. Geers and Firooz Aflatouni[*]

Department of Electrical and Systems Engineering, University of Pennsylvania, Philadelphia, PA 19104

[†]These authors contributed equally to this work.

[*]firooz@seas.upenn.edu

## Abstract

**Energy-efficient high-bandwidth interconnects play a key role in computing systems. Advances in silicon photonic electro-optic modulators and wavelength selective components have enabled the utilization of wavelength-division-multiplexing (WDM) in integrated optical transceivers, offering a high data-rate operation while achieving enhanced energy efficiency, bandwidth density, scalability, and the reach required for data-centers. Here, we report the demonstration of a single chip optical WDM PAM4 receiver, where by co-integration of a 32-channel optical demultiplexer (O-DeMux) with autonomous wavelength tuning and locking at a near-zero power consumption and a 32-channel ultra-low power concurrent electrical detection system, a record chip energy efficiency of under 0.38 pJ/bit is measured. The implemented 32 channel monolithic WDM optical receiver chip achieves an end-to-end latency of under 100 ps and a bit-error-rate of less than $10^{-12}$ with no equalization, predistortion, or digital-signal-processing, while operating at 1.024 Tb/s aggregate data-rate on a single input fiber, the largest reported data-rate for a WDM PAM4 receiver chip to date. The receiver bandwidth density of more than 3.55 Tb/s/mm$^2$ corresponds to more than an order-of-magnitude larger bandwidth density-energy efficiency product compared to the state-of-the-art optical PAM4 receivers for beyond 100Gb/s links. The chip, integrated using GlobalFoundries 45CLO CMOS-photonic process, can be used for implementation of energy-efficient high data-rate optical links for AI applications.**

## Introduction

The rapidly increasing bandwidth demands for different applications from computation to communication have necessitated revolutionary high speed and power efficient data movement solutions. Due to the skin effect and energy loss from electromagnetic radiation, high frequency signals suffer from greater attenuation in electrical interconnects. Consequently, as the data-rate increases, complex equalization techniques will be needed to compensate for the excessive channel loss leading to decreased power efficiency. On the other hand, optical links offer a high bandwidth, low propagation loss, and robustness against electromagnetic interferences, making them a good candidate for realization of high speed and long-reach interconnects and enabling the industrial deployment of 400G optical transceiver (TRX) modules [1, 2], as well as demonstrations of optical transceivers with near or beyond one tera-bit per second data-rates [3-6].

Silicon photonics technology has played a key role in the advancement of high-speed optical interconnects. The dense integration of on-chip optical transceiver modules further benefits multiplexing techniques such as wavelength-division multiplexing (WDM) or mode-division multiplexing (MDM), where the aggregate link data-rate is significantly enhanced [7-10].

A WDM optical transmitter (OTX) or optical receiver (ORX) can be implemented through hybrid or monolithic integration approaches, where in the former, the photonic integrated circuit (PIC) and electrical integrated circuit (EIC) are implemented in different processes and packaged using wire bonding [11-14] or flip-chip bonding [15-17] and in the latter, the photonic and electronic devices are fabricated on the same chip [18-21] leveraging the manufacturing infrastructure of the well-established CMOS [22, 23] or BiCMOS [24] technologies, forming an electro-photonic integrated circuit (EPIC). While the hybrid integration approach allows for utilization of state-of-the-art PIC and EIC, the effect of hybrid packaging crosstalk and parasitics must be carefully mitigated which could result in limited data-rate per wavelength and reduced energy efficiency. In contrast, the seamless, micron-scale interface between the photonic and electronic devices on a monolithic platform minimizes the parasitics and crosstalk, enabling the implementation of a fully integrated, single-chip, energy and area efficient solution for optical WDM links.

The monolithic WDM approach also presents certain design challenges. The shared substrate and close proximity between the high-speed electronics and the temperature-sensitive photonics can result in an undesired response shift in photonic devices due to the thermal crosstalk. While various on-chip wavelength stabilization techniques have been reported [14, 25-27], they are often based on thermal tuning of optical devices which reduces the overall link energy efficiency and could introduce its own thermal crosstalk challenges.

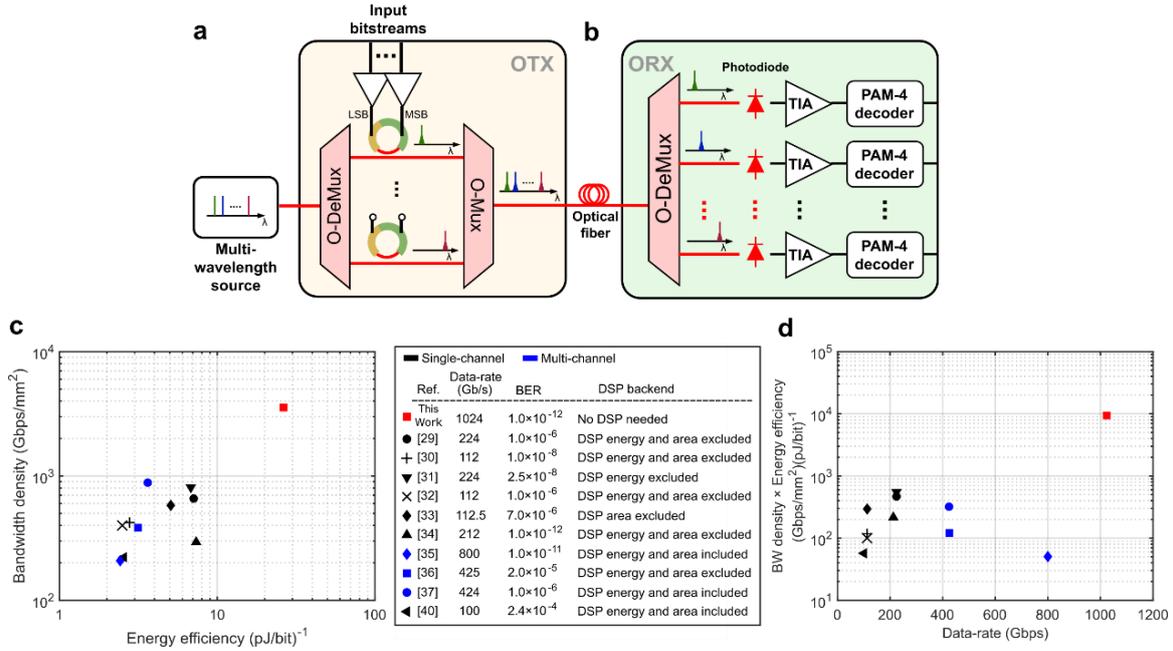

**Fig. 1 | Optical links using four-level pulse amplitude modulation. a,** Block diagram of a typical optical WDM PAM-4 transmitter. **b,** Block diagram of a typical optical WDM PAM-4 receiver. **c,** A survey of bandwidth density vs. energy efficiency for existing integrated CMOS PAM4 receivers with an aggregate data-rate larger than 100Gb/s. **d,** A survey of bandwidth density-energy efficiency product vs. aggregate data-rate for existing integrated CMOS PAM4 receivers with an aggregate data-rate larger than 100Gb/s. For bandwidth density, only core areas are considered (if not explicitly reported, the core area is estimated based on chip micrograph). WDM: wavelength-division multiplexing, OTX: optical transmitter, ORX: optical receiver, PAM4: four-level pulse amplitude modulation, O-Mux: optical wavelength multiplexer, O-DeMux: optical wavelength demultiplexer, TIA: trans-impedance amplifier. CMOS: complementary metal-oxide semiconductor. BER: bit error rate.

Utilizing four-level pulse amplitude modulation (PAM4) [25, 26, 28, 29] instead of non-return-to-zero (NRZ) can further improve the bandwidth of optical and wireline links by relaxing the bandwidth requirements for the ORX front-end and reducing the complexity of the clock generation and distribution at the cost of reduced link SNR and an increased linearity requirement for the PAM4 ORX front-end.

End-to-end PAM4 receivers often utilize equalization and digital signal processing (DSP) backend to improve the detection bit-error-rate (BER) [29-34], which could result in lower system energy efficiency and bandwidth density. Furthermore, to increase the aggregate data-rate, PAM4 systems are typically implemented as multi-channel modular systems with one fiber per channel [35-37], which despite excellent performance, results in a higher packaging complexity and cost. The number of optical fibers, and hence the packaging and routing complexity, could be reduced through WDM approach. A WDM approach with a large number of carriers could also improve the overall energy efficiency and bandwidth density. For a target aggregate data-rate, using a larger number of optical carriers in a WDM scheme allows for a lower per-carrier data-rate, which could relax or eliminate the need for equalization and backend DSP, significantly increasing the system energy efficiency and bandwidth density.

A WDM PAM4 optical link consists of a multi-wavelength light source (e.g. an optical frequency comb source or a laser array) a PAM-4 OTX and a PAM-4 ORX. Fig. 1a shows the block diagram of the transmitter side of a link. A multi-wavelength light source is coupled into the PAM4 OTX, where the carriers are separated using an optical demultiplexer (O-DeMux), individually modulated (in this figure, as an example, using multi-section ring modulators [38] driven by input bitstreams) and multiplexed to form the OTX output. A single mode optical fiber is used to route the OTX output to the input of the WDM PAM-4 ORX (Fig. 1b), where the modulated carriers are demultiplexed using an O-DeMux and individually detected and demodulated using PAM4 decoders.

Here we report the demonstration of a single monolithic WDM PAM4 optical receiver chip that achieves an error-free operation at record data-rate of 1.024 Tb/s on a single input fiber with an energy efficiency of under 0.38 pJ/bit and a bandwidth density of 3.5 Tb/s/mm$^2$. No equalization or digital signal processing (DSP) were used in the reported system, resulting in a significant reduction in energy consumption and chip area. The chip, implemented on GlobalFoundries 45CLO CMOS-photonics process consists of a 32-channel O-DeMux and a 32-channel ultra-low power detection and PAM-4 decoding system array. The optical multiplexer utilizes capacitive phase shifters to perform autonomous wavelength tuning and locking at a near-zero power consumption [38, 39]. Each output of the O-DeMux is detected using a silicon-germanium photodiode and the resulting photocurrent is amplified using a trans-impedance amplifier and decoded. On-chip per-channel de-serializer and BER measurement systems are used to ease the multi-channel BER measurements. The 32 channel ORX chip operates in the optical C-band at a per-wavelength data-rate of 32 Gb/s and was integrated within a footprint of 4.2 mm$^2$, where each ORX channel, including the de-serializer and BER analyzing back-end, occupies an active area of 0.07 mm$^2$.

Fig. 1c and Fig. 1d present surveys of the existing CMOS PAM4 ORX systems and provide a comparison between this work and the state-of-the-art in terms of bandwidth density versus energy efficiency and bandwidth density-energy efficiency product versus the aggregate data-rate [29-37, 40], where, compared to the state-of-the-art reported end-to-end CMOS PAM-4 receivers with an aggregate data-rate higher than 100 Gb/s, the implemented chip achieves a record energy efficiency (more than 5 times compared to the reported end-to-end PAM4 ORX) and more than an order-of-magnitude higher bandwidth density-energy efficiency product, while achieving a record aggregate data-rate of 1.024 Tb/s on a single fiber, enabling the realization of low-cost energy-efficient optical interconnects with ever-growing applications in data centers and AI systems.

**Results**

**Optical demultiplexer with autonomous tracking of carriers**

The 32 PAM4 modulated carriers with 200 GHz carrier-to-carrier spacing in the optical C band are delivered to the implemented PAM4 ORX chip using a single-mode fiber, coupled into the chip through the input grating coupler (with a coupling loss of about 5 dB) and routed to the input of the 1:32 O-DeMux using the input nanophotonic waveguide, where the carriers are demultiplexed and routed for detection and PAM4 demodulation.

The monolithic integration on a CMOS process allows for implementation of capacitive optical phase shifters with zero static power consumption [38, 39] alongside the tuning control electronics. Fig. 2a shows the cross-section of the capacitive phase shifters implemented in the GlobalFoundries GF45CLO process, where the polysilicon layer primarily used in the gate of CMOS transistors is repurposed to form a polysilicon-oxide-silicon optical waveguide structure. In this case, the vertically stacked doped silicon and polysilicon layers, separated by a thin $SiO_2$ layer also forms a capacitor. The thin oxide layer does not directly interact with the optical mode but blocks the DC current when a DC voltage is applied across the capacitor. While this configuration consumes no static energy, the applied voltage modifies the charge distribution in the lightly doped waveguide region, where the overlap between the voltage-dependent charge profile and the optical mode within the waveguide alters the effective refractive index and induces an optical phase shift across the capacitive phase shifter. Fig. 2b and 2c show the structure of the ring resonator and MZI, which serve as the core components of the DeMux system. Both the MZI and the ring resonators employ a hybrid configuration of capacitive phase shifters and heaters for wavelength tuning.

Fig. 2d shows the block diagram of the 1:32 O-DeMux, which consists of a 1:8 Mach-Zehnder interferometer (MZI) binary tree stage followed by an 8:32 drop-ring resonator array. The performance of the O-DeMux system is determined by different factors such as total insertion loss, channel-to-channel isolation, and the required energy for wavelength alignment and carrier tracking.

A carrier spacing of 200 GHz is used to limit inter-channel crosstalk at the target data rate of 32 Gb/s/channel. However, due to the finite quality factors typically achievable in capacitively tuned micro-ring resonators [39], this spacing may still yield non-negligible crosstalk during wavelength demultiplexing. To address this, a 1:8 Mach-Zehnder interferometer (MZI) binary tree is employed at the frontend to effectively increase the channel spacing to 1600 GHz prior to spectral filtering by the ring resonators.

For 32 carriers with wavelength of $\lambda_1$ to $\lambda_{32}$, the length of the delay imbalance between the arms of the MZIs in the first, second, and third stages of the 1:8 MZI binary tree structure is set to 180 μm, 95 μm, and 47.5 μm, corresponding to FSRs of 400, 800, and 1600 GHz, respectively. As a result, the 1:8 MZI binary tree O-DeMux separates the 32 input carriers into eight batches of four carriers with a carrier spacing of 1600 GHz for each batch (i.e. the $i^{th}$ batch consists of carriers with wavelengths of $\lambda_i$, $\lambda_{i+8}$, $\lambda_{i+16}$, and $\lambda_{i+24}$ for i = 1,2,…,8).

The 8:32 O-DeMux consists of eight branches. On each branch, three ring resonators are placed on the main bus waveguide. Each ring resonator selects one of four carriers at its drop port with the 4th carrier remaining on the bus waveguide.

Ideally, the FSR of the ring resonators should be large enough to only select the target carrier without overlapping with other carriers. However, in practice, the FSR of the ring resonators is limited by the smallest practical size of the ring set by the fabrication constraints and the quality factor deterioration. As a result, careful frequency planning and FSR design is needed as the presence of multiple undesired in-band resonances could result in excess crosstalk. Additionally, there is a trade-off between the drop-port insertion loss and the crosstalk as a function of the ring coupling coefficient, which should be considered in the ring resonator design. In the implemented O-DeMux, the FSR of the ring resonators were designed to be between 707 GHz and 713 GHz (by setting by slight difference in the circumference of the ring resonators). More details on the design parameters and trade-offs is included in the Methods section and the Extended Data Fig. 1.

To compensate for the effect of fabrication process variations and temperature fluctuations, capacitive optical phase shifters are integrated within MZI/ring devices that can tune their optical response at a zero static power consumption significantly improving the link energy efficiency. While effective, the capacitive modulator tuning range is limited by the capacitor breakdown voltage and/or the loss introduced due to the increase in the imaginary part of the refractive index. Typically, the capacitive tuning is sufficient to compensate for the effect of process variations, however, to ensure uninterrupted operation of the link for the cases that a larger tuning range is needed, besides the capacitive section, a thermal phase shifter is also place within MZI/ring devices. In our design, to reduce the O-DeMux insertion loss (induced by the loss of the capacitive section, thermal phase shifters are utilized in the first (1:2 DeMux) and second (2:4 DeMux) stages (total of three MZIs), while for the third MZI layer (consists of four MZIs) and all rings a hybrid capacitive/thermal phase shifter system is employed (although the capacitive tuning is almost always sufficient). This approach optimizes both optical insertion loss while significantly reducing the total tuning power consumption. Each phase shifter within an MZI or ring resonator device is electronically controlled using a sensing, actuation, and memory.

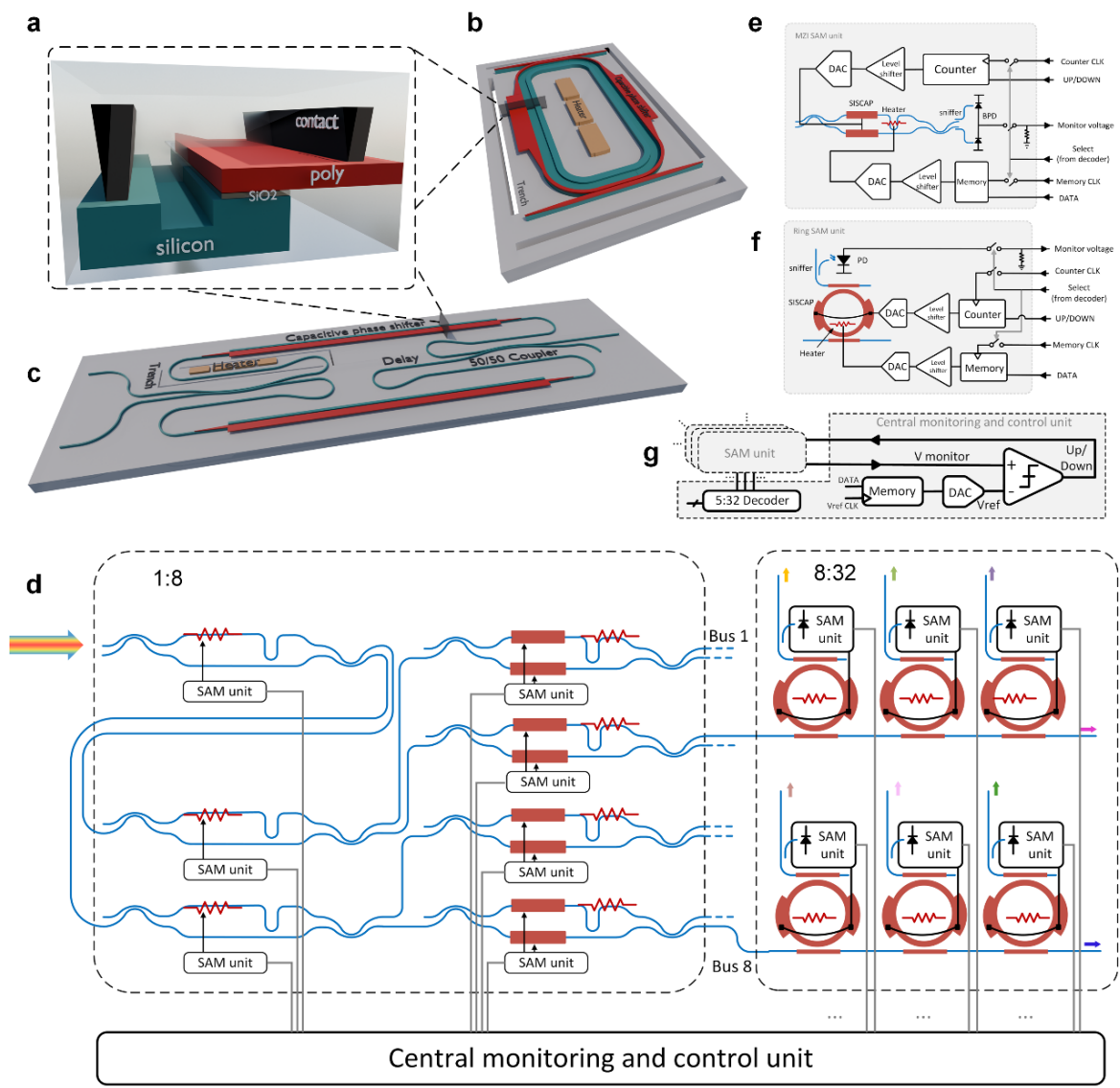

**Fig. 2 | MZI and ring resonator hybrid O-DeMux block diagram. a,** Cross-section of the capacitive

phase modulator implemented on GF45CLO process. **b,** Capacitively tuned drop ring resonator. Heaters are added in case of an insufficient capacitive tuning range. Trenches are used around the ring resonator to improve the thermal phase shift efficiency. **c,** Capacitively tuned MZI with broadband couplers at the input and output. **d,** Block diagram of the implemented 1:32 demultiplexer based on a 1:8 MZI binary tree followed by a 8:32 ring resonator array. Block diagrams of **e,** the MZI control unit (SAM) and **f,** the ring resonator control unit (SAM). **g,** Schematic of the central monitoring and control unit that sequentially selects the SAM units to be engaged in the carrier tracking and locking system.

(SAM) unit placed adjacent to each device (Fig. 2e-f). The SAM units are sequentially selected and activated (using a global on-chip decoder) and used in a feedback loop (as shown in Fig. 2g) to wavelength lock each MZI or ring resonator to the corresponding carrier. For the SAM unit within each MZI, 2% of the output of the MZI output power is tapped off and converted to a voltage and monitored. This monitor voltage is compared with a stored reference value to generate an UP/DOWN signal for a 7-bit counter. The counter output is converted to an analog voltage and is used to drive the phase shifter.

The balanced capacitively tuned MZIs in Fig. 2d utilizes a 250 μm long capacitive section on each arm, which could be driven differentially. Fig. 3a shows the capacitive tuning response and power consumption for a single balanced MZI within the 1:8 O-DeMux, where by applying a differential voltage across the two capacitive sections, a tuning range of about 0.5 free-spectral range (FSR) is achieved (a 0.22 FSR blueshift and a 0.28 FSR redshift), while consuming zero static power. Measurements show that this capacitive tuning range is mostly sufficient to align the MZI to the target wavelength grid, however, thermal phase shifters were also added for the unlikely cases that the tuning range needs to be further increased. Note that, while consuming zero static power, the capacitive phase shifters typically introduce more optical loss than thermal phase shifters [39], therefore, to decreases the overall optical loss, the capacitive sections in the first three MZIs of this design (one in the first layer and two in the second layer) are removed and these MZIs can only be thermally tuned. The MZI thermal tuning is performed using an N-doped silicon resistor, serving as a heater, placed next to the waveguide of the bottom arm. Undercut deep trenches are placed around the heater (Fig. 2c) to increase the thermal isolation and improve the thermal tuning power consumption while reducing the thermal leakage and crosstalk. Fig. 3b shows the thermal tuning response and power consumption for a single MZI. The MZI can be thermally tuned by more than an FSR. The power needed to introduce a $\pi$ radians optical phase shift across the thermal phase shifter, $P_\pi$, is approximately 10 mW. Fig. 3c shows an example wavelength locking process for the MZIs within the 1:8 MZI binary tree. First, during the acquisition process for each MZI, the 6-bit counter within the SAM unit is used to tune the wavelength of the notch response of MZI while the MZI response relative to the wavelength of the target carrier is monitored using the balanced photodiode in Fig. 2e, whose output photocurrent is converted to the monitor voltage using a resistor. Once the MZI is tuned across the entire tuning range, the reference value is set the largest recorded monitor voltage. During the wavelength locking process, the counter tunes the MZI until the monitor voltage reaches the reference value in which case, the feedback loop dynamically adjusts the counter UP/DOWN signal to ensure the MZI is locked to the wavelength of the target carrier by keeping the monitored voltage the same as the reference value. Once the wavelength alignment of all MZI devices is complete, the three-layer MZI binary tree (the 1:8 O-DeMux) transfer function will be aligned to the wavelength grid of the input carriers as shown in Fig. 3d.

Ring resonators within the 8:32 O-DeMux can also be capacitively or thermally tuned. As shown in Fig. 2b, a capacitive section placed along the straight section of the ring resonators ring resonators allows for capacitive tuning of its resonance wavelength at zero static power consumption. Fig. 3e shows the ring resonator capacitive wavelength tuning response and power consumption, where a maximum blueshift of 0.72 nm in the resonance wavelength is measured. For a case where the capacitive tuning range is insufficient, thermal tuning can be utilized using an N-doped silicon structure (serving as a heater), which is placed at the center of the ring resonator. Undercut trenches are placed around the ring resonator structure to increase thermo-optic modulation efficiency and reduce the thermal crosstalk. Fig. 3f shows the ring resonator thermal wavelength tuning response and power consumption, where a power efficiency of 22 mW/nm is measured. Fig. 3g shows an example wavelength locking process for the ring resonators within the 8:32 O-DeMux, where similar to the wavelength locking process for MZIs, first, an open-loop wavelength sweeping (using the 5-bit counter) is performed and the maximum monitor voltage is recorded as the reference value and then, during the wavelength locking phase, the ring resonance frequency is locked to the wavelength of the target carrier.

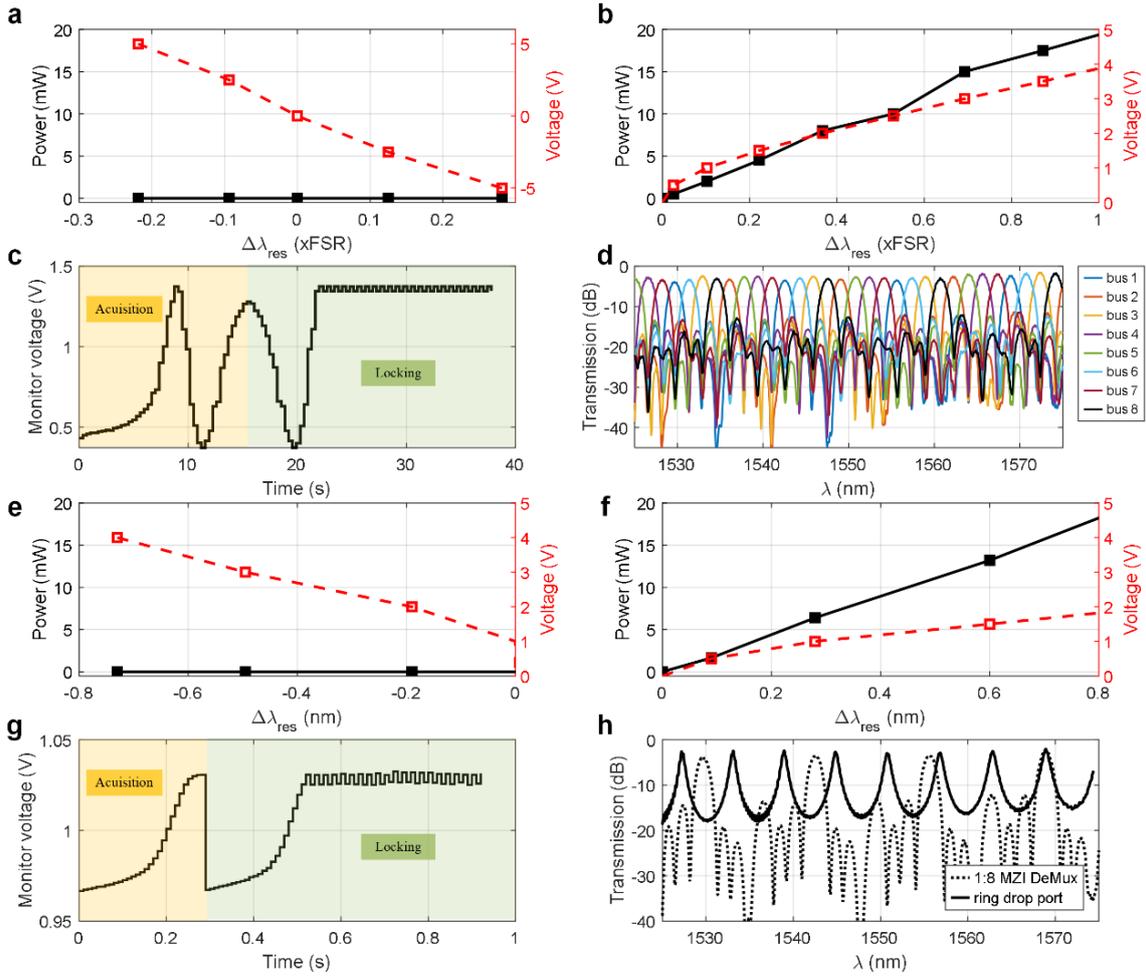

**Fig. 3 | MZI and ring based DeMux alignment. a,** MZI capacitive wavelength tuning response and power consumption. **b,** MZI thermal wavelength tuning response and power consumption. **c,** Measured wavelength locking process for the 1:8 MZI based O-DeMux. First, during the open-loop (acquisition) phase, the phase shifter voltage is swept (using the 6-bit counter) and the largest detected monitor voltage is recorded as the reference value. Then, during the closed loop (locking) phase, the MZI is locked to the target wavelength. The process is slowed down to ease the illustration. **d,** Transfer function of the 1:8 MZI based O-DeMux after wavelength alignment. **e,** Ring resonator capacitive wavelength tuning response and power consumption. **f,** Ring resonator thermal wavelength tuning response and power consumption. **g,** Demonstration of the capacitive phase locking for a ring resonator including an open-loop wavelength sweeping (using the 5-bit counter), where the maximum monitor voltage is acquired and recorded as the reference value, followed by the locking of the ring resonator the target wavelength. **h,** Transfer function of drop port of a single ring resonator placed after the wavelength-aligned 1:8 MZI O-DeMux stage, where the ring is tuned to one of the carriers at 1569nm.

After wavelength alignment, the 1:32 O-DeMux system achieves an insertion loss of 4–5 dB, with 1–2 dB attributed to the 1:8 MZI tree stage (Fig. 3. (d)) and the remainder resulting from the drop port transmission of the ring resonators. Fig. 3. (h) shows the transfer function of the drop port of a single ring resonator placed after a wavelength-aligned 1:8 MZI O-DeMux stage. The ring is wavelength-tuned to align with one of four incoming wavelength channels at 1569 nm.

The spectral isolation between WDM channels is determined by the wavelength mapping, the MZI rejection ratio, and the full width at half-maximum (FWHM) bandwidth of the ring resonators. The MZI rejection ratio is primarily determined by the quality of its input and output 50-50 directional couplers. With broadband 50-50 directional couplers, the MZIs achieve a rejection ratio exceeding 25 dB across a 50 nm bandwidth. Consequently, the overall maximum crosstalk is constrained by the isolation provided by the ring resonators, which is determined by multiple ring resonator parameters including FSR, propagation loss, and coupling coefficient. More details on the design of the broadband directional couplers are included in the Methods section and Extended Data Fig. 2.

The measured total insertion loss and the average energy efficiency of the 1:32 O-DeMux at 1.024 Tb/s is 4 dB and 8 fJ/bit, respectively, with a channel-to-channel crosstalk between -25 dB and -10 dB. More details on the calculation of required tuning power consumption are included in the Methods section and Extended Data Fig. 3.

**PAM4 detection and decoding**

Exploiting the very small parasitic components of the integrated photodiodes and micron-scale metal routing in our monolithic approach, the design objective of the PAM4 detection and decoding (DE) system is to support error-free operation without equalization to substantially reduce the power consumption and design complexity. Additionally, an inductor-less architecture could further reduce the chip area and improve the areal bandwidth density.

The block diagram of the implemented PAM4 detection system is shown in Fig. 4a. Each of the 32 outputs of the O-DeMux is photo-detected and amplified using a multi-stage trans-impedance amplifier (TIA). The outputs of the TIAs are 4-to-1 multiplexed and routed to PAM4 decoders followed by de-serializer and BER measurement system blocks enabling on-chip measurements of all channels, eight at a time, while reducing the required chip area.

Fig. 4b shows the schematic of the TIA. To achieve a differential operation, the input light is equally split using a Y-junction and each output of the Y-junction is photo-detected using a reverse biased silicon-germanium photodiode (PD). The anode of first (top) PD is grounded while its cathode is connected to the 1.5V supply through a 850 $\Omega$ resistor and the cathode of the second (bottom) PD is grounded and its anode it connected to -1.5 V supply through another 850 $\Omega$ resistor. The differential detection voltage is generated between the cathode of the first PD and the anode of the second PD.

This differential voltage is amplified using a differential source follower (DSF) stage with a cross-coupled n-channel metal oxide field effect transistor (MOSFET) pair, which provides gain while enhancing the bandwidth. The DSF outputs are buffered using two source follower stages and routed to the PAM4 decoder input. More details on the TIA design are included in the Methods section.

The PAM4 decoder used in the implemented receiver is a 2-bit time-interleaved flash analog-to-digital converter (ADC) that converts the 4-level analog signal detected by the TIA into two digital bit streams representing the most significant bits (MSB) and least significant bits (LSB) of each detected symbol. Fig. 4c shows the schematic of the PAM4 decoder, which consists of two half-rate quantizers. Each quantizer includes three slicers for threshold comparison and a circuitry for thermometer to binary code conversion. More details on the implementation of the slicers is included in the Methods section and Extended Data Fig. 4. The de-serializer utilizes three layers of D-type flip flops (DFF) for the 2-to-4, 4-to-8 de-serializing and the output data alignment, respectively. Each on-chip bit-error-rate (BER) measurement system includes an 8-way parallel pseudo-random binary sequence 7 (PRBS7) generator [41], which is triggered by an off-chip reset (RST) signal to first take initial pattern from the de-serializer output and then generate input-independent data as the target pattern. The details of the de-serializer and bit-error-rate measurement system are included in the Method section and Extended Data Fig. 4.

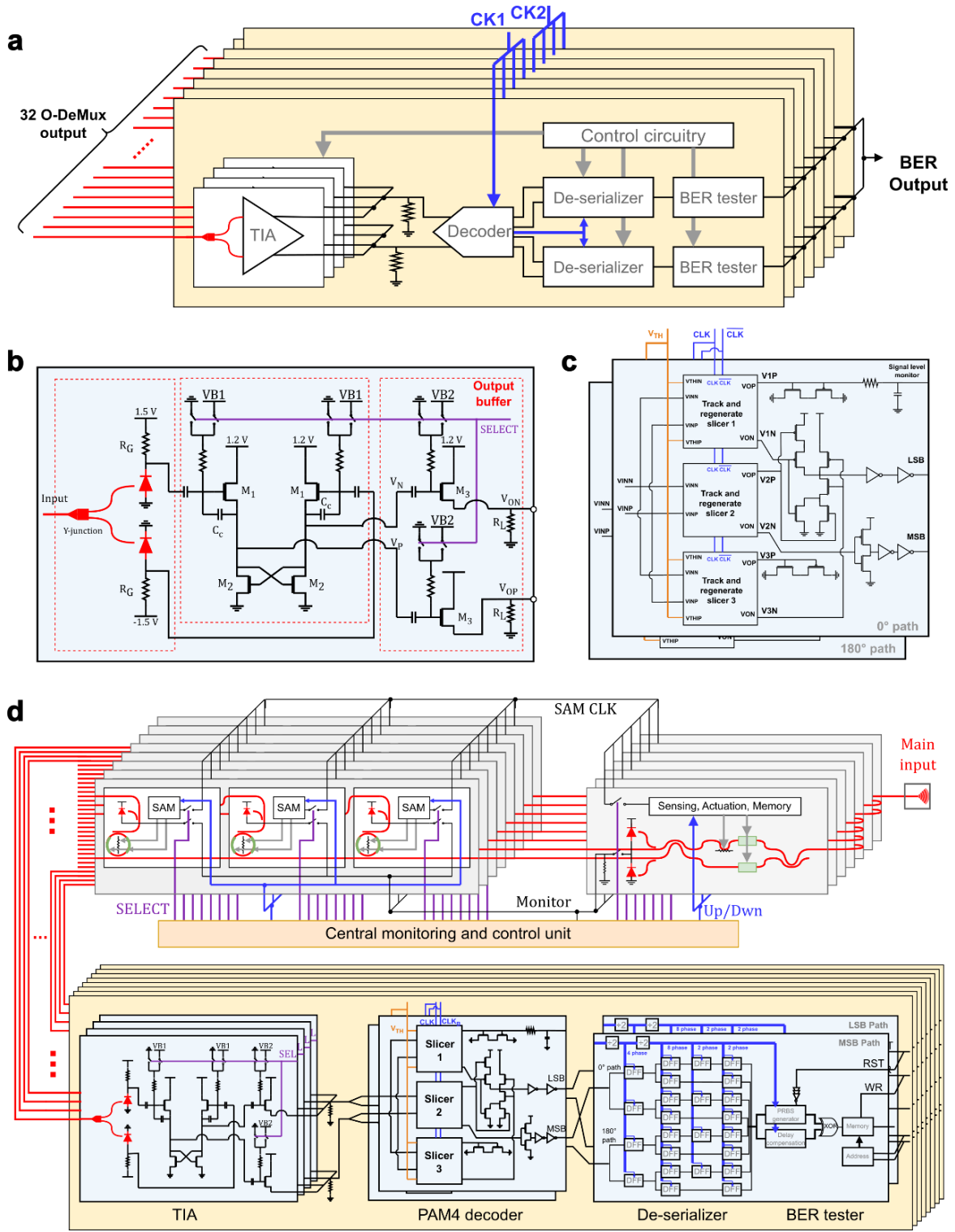

**Fig. 4 | Receiver electronics and system implementation. a,** Schematic of the implemented PAM4 detection system of the implemented Optical receiver. The 32 output waveguides of the O-DeMux are routed to 32 photoreceivers composed of dual photodiodes and differential trans-impedance amplifiers (TIA) and detected. The outputs of the TIAs are 4-to-1 multiplexed and routed to PAM4 decoders and subsequent de-serializer and BER measurement blocks. **b,** Schematic of the differential TIA and buffer. **c,** Schematic of the PAM4 decoder including 3 track and regenerate slicers. **d,** Diagram of the implemented 32 channel WDM ORX including the 1:32 O-DeMux with per-channel autonomous

wavelength locking and tracking, followed by 32 differential photoreceiver, an array of PAM4 decoders, de-serializers, and on-chip bit-error-rate measurement systems (i.e. BER tester).

**System integration**

Fig. 4d shows the diagram of the implemented 32 channel PAM4 WDM optical receiver system, which includes the 1:32 O-DeMux with per-channel autonomous wavelength locking and tracking, followed by 32 differential photoreceiver, an array of PAM4 decoders, de-serializers, and on-chip bit-error-rate measurement systems.

The optical receiver in Fig. 4d was integrated using GlobalFoundries monolithic 45nm CMOS-silicon photonics process (45SPCLO). Fig. 5a shows the layout of the 32-channel WDM ORX chip, where the electronics are implemented as 8 identical modules with each module occupying only 0.1296 mm$^2$. Fig. 5b shows the micrograph of the 32-channel optical receiver chip implemented within a footprint of 4.72 mm$^2$.

To fully characterize different blocks of the system a single-channel ORX with on-chip de-serializer and BER measurement system was fabricated as the test chip. The Micrograph of the test chip, with a core area of 0.07 mm$^2$, is shown in Fig. 5c. Both the main and the test chips were in-house wire-bonded to custom PCBs for measurements.

Figure 5d shows the block diagram of the setup used to measure the performance of the single channel test chip. A laser emitting 5 dBm at 1550 nm is polarization adjusted using a polarization controller and modulated using a Fujitsu FTM7937EZ modulator. The modulator output is amplified using an erbium-doped fiber amplifier (EDFA) to about 8 dBm and is coupled to the chip using a single mode fiber. One channel of 100 GS/s arbitrary waveform generator (AWG) system (Micram DAC10002) is used to generate the electrical PAM4 / NRZ waveforms, which is used to drive the modulator. The AWG also generates a clock signal that is synchronized to the data stream with an adjustable delay. The clock frequency, which is the same as the baud-rate of the input data stream, is divided by 2 using an error analyzer (EA 40 GIG, SHF), amplified, converted to a differential signal (using a 180º RF hybrid), DC-shifted (using bias tees) and used to serve as the clock signal for the ORX test chip. Bias voltages of the modulator are carefully set for a linear response resulting in equally spaced PAM4 levels. The modulated optical signal is delivered to the ORX test chip through a single-mode fiber using the on-chip grating coupler with an approximate coupling loss of 5 dB. Uncorrelated PRBS7 patterns are combined to generate the PAM4 data. The measured PAM4 eye diagram at 32 Gb/s at the TIA output within the ORX test chip (before the PAM4 decoder) is shown in Fig. 5e. The ORX test chip can be configured to operate as an NRZ receiver. Fig. 5f shows the measured NRZ eye diagram at 24 Gb/s at the TIA output.

To measure the BER, the optical modulation amplitude (OMA) and sampling offset (between the clock and data signals) of the ORX test chip for both 32 Gb/s PAM4 and 24 Gb/s NRZ are varied while the on-chip BER tester is used to measure the chip BER. Fig. 5g shows the measured BER versus OMA for the ORX test chip for both 32 Gb/s PAM4 and 24 Gb/s NRZ formats. For this measurement, the average optical power is measured after the grating coupler using a sniffer, which taps 5% of the signal into a monitoring PD for average current detection and the OMA is calculated based on the modulator ER and the calculated average optical power. In the PAM4 mode, BER of both the decoded MSB and LSB bitstreams are measured. Since the LSB results are decoded using all three slicers (in Fig. 4c) while for the MSB decoding, only the middle slicer is used. As a result, the measured BER of MSB signal is always lower than that of LSB bitstream. Therefore, only the measured BER of the LSB in the PAM4 mode is included in Fig. 5 to simplify the visualization. In the NRZ mode, MSB signal is decoded using the middle slicer while the top and bottom slicers are disengaged (by setting the decoder threshold voltages close to power supply rails).

The measured BER for the PAM4 and NRZ formats versus sampling offset (the bathtub graphs), for the ORX test chip are shown in Figs. 5h and 5i respectively.

The same measurement setup shown in Fig. 5d is used to characterize the 32-channel WDM PAM4 ORX chip. To perform the chip characterization, first, the on-chip optical DeMux is tuned to the input wavelength grid (using a pre-extracted channel mapping look-up table) optimizing the power delivered to each PD while minimizing channel-to-channel crosstalk. This is accomplished by sequentially selecting the SAM control units of MZI/ring devices, reading its operation point, and if needed, adjusting the count number within the selected SAM unit to maintain wavelength alignment of the MZIs and ring devices to the corresponding optical carriers.

Figure 5j shows the measured BER as a function of OMA for a selected channel of the 32-channel WDM OAM4 receiver, where the chip achieves a BER of $10^{-12}$ for an input OMA higher than -3.0 dBm and -8.3 dBm for 32 Gb/s PAM4 and 24 Gb/s NRZ modes, respectively, which is in close agreement with the measurement results of the ORX test chip. The measured bathtub curves for the 32-channel receiver chip for 32Gb/s PAM4 and 24Gb/s NRZ modes are shown in Fig. 5k and 5l, respectively, where at the reported sensitivity levels, the 32-channel ORX chip achieves larger than 0.05 UI opening for a BER of $10^{-12}$ for both 32 Gb/s PAM4 and 24 Gb/s NRZ modes.

All 32 channels of the WDM PAM4 receiver chip were tested using PAM4 modulated carriers. The channel-wavelength mapping diagram is shown in Fig. 5m and the measured BER of all channels is shown in Fig. 5n, where a BER of <$10^{-12}$ for all 32 channels at 32 Gb/s per channel (an aggregate data rate of 1.024 Tb/s) was achieved.

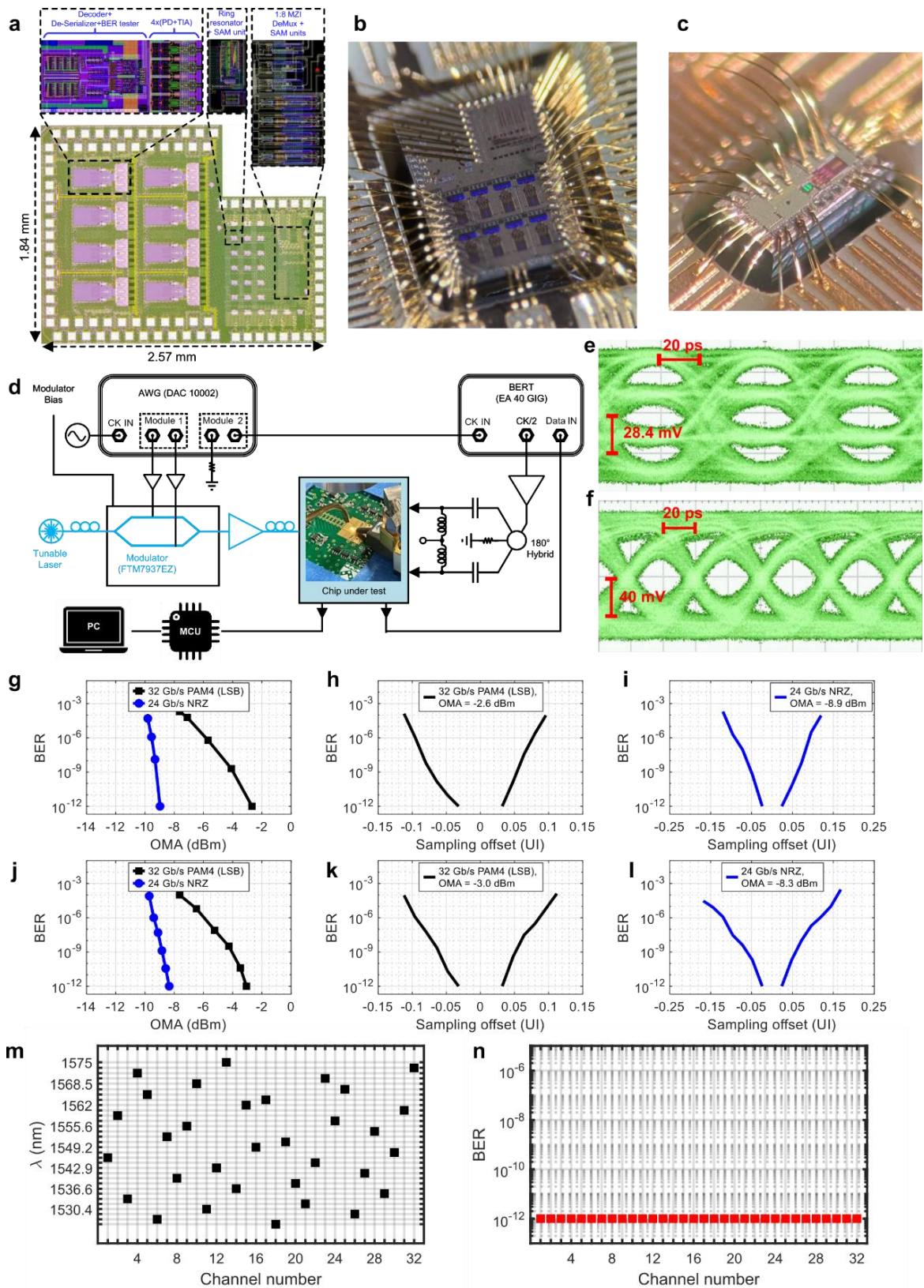

**Fig. 5 | Measurement results for WDM PAM4 ORX. a,** Layout of the 32-channel WDM PAM4 ORX chip with zoomed-in key blocks. **b,** The micrograph of the 32-channel WDM PAM4 ORX chip. **c,** Micrograph of the test chip, a single channel of the ORX including the BER measurement system. **d,** The measurement setup for characterization of the test chip and the 32-channel WDM PAM4 ORX chip.

**e,** Measured PAM4 eye diagram at 32 Gb/s at the TIA output in the test chip. **f,** Measured eye diagram at the TIA output of test chip operating in the NRZ mode at 24 Gb/s. Measured ORX **g,** sensitivity and bathtub curves for **h,** 32 Gb/s PAM4 operation and **i,** 24 Gb/s NRZ operation using the test chip. Measured ORX **j,** sensitivity and bathtub curves for **k,** 32 Gb/s PAM4 operation and **l,** 24 Gb/s NRZ operation for the 32-channel chip. **m,** The wavelength-channel mapping diagram. **n,** The measured PAM4 BER (LSB) for all channels operating at 32Gb/s (1Tb/s aggregate data-rate).

**Discussion**

In this work, we have demonstrated a 32-channel WDM PAM4 receiver chip that uses only a single mode fiber at its input and operated at a record data-rate of 1.024 Tb/s with a BER less than $10^{-12}$. The measured TIA power consumption, including PD bias, is 6.89 mW and the measured PAM4 decoder power consumption, including the clock distribution, is 4.88 mW, resulting in an overall energy efficiency of 0.37 pJ/bit for the PD and the electronics, which considering the 8.5 fJ/bit energy efficiency of the O-DeMux, results in a total energy efficiency of about 0.38 pJ/bit for the implemented 32-channel WDM PAM4 receiver chip, which is more than five times higher than the stat-of-the-art end-to-end CMOS PAM4 receivers with an aggregate data-rate larger than 100Gb/s. More details on the energy efficiency calculations for the O-DeMux are included in the Methods section. The design achieves a bandwidth density of more than 3.55 Tb/s/mm$^2$ which results in bandwidth density-energy efficiency product which is more than an order-of-magnitude higher than that of the state-of-the-art PAM4 receivers with a data-rate larger than 100Gb/s. This significant improvement is a result of the small integrated PD capacitance (~10 fF) and seamless on-chip photonics-to-electronics interconnects, as well as an equalizer-free design. Furthermore, compared to typical multi-channel designs, where multiple optical fibers are used to deliver the data to the system, the reported chip utilizes a single fiber with 1.024 Tb/s/fiber density, relaxing the packaging (e.g. fiber attachment) complexity and cost.

More details on the ring resonator design are included in the Methods section and the Extended Data Fig. 1. The design of the broadband directional coupler used within the MZI devises is discussed in the Methods section and Extended Data Fig. 2. The statistical analysis on the power consumption of the O-DeMux is included in the Methods section and Extended Data Fig. 3. The detection electronic circuits are further discussed in the Methods section and Extended Data Fig. 4. The performance of the implemented 1.024 Tb/s WDM PAM4 receiver is compared with selected reported multi-carrier PAM4 receivers in the Extended Data Table 1.

**Methods**

**Chip fabrication**

The ORX system was monolithically integrated using the GlobalFoundries 45CLO CMOS-SOI silicon photonic platform. This technology supports devices with an $f_T$ of up to 280 GHz [23], making it well-suited for RF demodulation on the same chip alongside the optical DeMux system and corresponding electronics. The monolithic integration enables the realization of optical transceivers with reduced packaging complexity and less parasitics.

The design incorporates photonic waveguides with an optical loss of approximately 1.4 dB/cm at around 1550 nm, grating couplers with a coupling loss of about 4.5-5 dB, and photodiodes exhibiting a responsivity of approximately 0.9 A/W and bandwidth of about 50 GHz.

**Ring resonator design**

In the proposed O-DeMux architecture, three drop-ring resonators are placed on a common bus waveguide to demultiplex four wavelength channels spaced at 1600 GHz. Each ring is designed to select a carrier with a specific wavelength, while one (last) tone remains at the bus waveguide output. To minimize the channel-to-channel crosstalk in the ring-based 8:32 demultiplexer, the free spectral range (FSR) of the rings, determined by their effective circumference, must be carefully optimized, as extra resonances between consecutive carriers may degrade the O-DeMux performance. Extended Data Fig. 1a shows the simulated optical crosstalk as a function of FSR for various ring coupling coefficients, $k^2$, for a single ring used in one of 8 branches of the 8:32 ring resonator-based O-DeMux (i.e. the cascaded three ring configuration on a waveguide bus). Note that, here, optical crosstalk is defined as the ratio of the power coupled from non-target carriers and the power transferred from the target carrier. The results show that an optimal FSR of about 710 GHz minimizes crosstalk, Therefore, three ring resonators with slightly different circumferences are designed such that their FSRs lie within the 709 GHz to 711 GHz range.

Extended Data Fig. 1b illustrates the trade-off between drop-port insertion loss and crosstalk for a single ring resonator (with an FSR of 710 GHz) as a function of its coupling coefficient. In this work, a coupling coefficient of 0.3 was used to achieve insertion loss of less than 2 dB, resulting in a crosstalk of less than -10 dB.

**MZI broadband coupler design**

Each MZI in the O-DeMux system utilizes two broadband curved 50% directional couplers at the input and output to evenly split and recombine the light [42]. As shown in Extended Data Fig. 2a, these couplers consist of two bent single-mode waveguides optimized for peak performance near 1550 nm with design parameters of $R = 46.27\ \mu m, \theta = 9.41°$, and $gap = 0.2 \mu m$. Extended Data Fig. 2b shows the measured normalized transmission response of an unbalanced MZI incorporating these broadband couplers, demonstrating an extinction ratio exceeding 25 dB with a variation of approximately 1.5 dB across a 50 nm bandwidth.

**DeMux power consumption**

The O-DeMux system consists of three thermally tuned MZIs, four capacitively tuned MZIs (that can also be thermally tuned if needed), and 24 capacitively tuned micro-ring resonators, that can also be thermally tuned if needed. The analysis presented here is to estimate the required overall O-DeMux power consumption needed to keep it aligned with all 32 carriers in presence fabrication process variations.

To evaluate the impact of process variations in the GlobalFoundries 45CLO platform at 1550 nm, an array of 150 identical ring resonators was fabricated over a 1.4 mm × 0.9 mm area. Each capacitively tuned ring had a circumference of 96 µm and was formed using 0.5 µm-wide waveguides, resulting in an average FSR of 6.3 nm. A laser was coupled into the chip and optically split to distribute equal power to all resonators. The output from each resonator was photo-detected, and an electronic decoder was used to sequentially measure the response of each micro-ring devices. The laser wavelength was swept while monitoring all resonators. The measured resonance wavelength for all micro-ring devices and its distribution (with a Gaussian fit) are shown in Extended Data Fig. 3a and 3b, respectively, where a mean resonance wavelength of 1558.59 nm and a standard deviation of about 0.13 nm is observed. This variation range is within the capacitive tuning range of the ring resonators within the 8:32 O-DeMux (i.e. 0.7 nm) indicating that capacitive tuning with a zero power consumption is sufficient to compensate for the effect of process variations. To investigate the effect of processes variations on the MZIs within the O-DeMux system, the measurement results presented in Extended Data Fig. 3b were used to estimate the normalized standard deviation of the waveguide group index as $8.5 \times 10^{-5}$. Considering this estimate for the group index variance, 10,000 MZI devices were simulated. The result indicates that the capacitive tuning range of the MZI is sufficient to overcome the effect of process variations with zero power consumption. Similarly, assuming the effect of process variations as presented in Extended Data Fig. 3b, simulations show that the power consumption for the first three thermally tuned MZIs within the 1:8 O-DeMux is about 2 mW per MZI. With the digital control circuit consuming 0.3 mW of static power, the expected total tuning power consumption of the 1:32 O-DeMux system is about 6.3 mW corresponding to an energy efficiency of 6.3 fJ/bit at 1.024 Tb/s aggregate data-rate.

**Detection electronics**

Given the small photodiode capacitance in our monolithic integration approach, in the TIA schematic in Fig. 4b, a relatively large passive gain of 850 Ω is realized using resistors $R_G$ to significantly reduce the power consumption while, at the same time, by utilizing a differential source follower (DSF) structure the bandwidth and noise performance requirements are met. The output stage of the TIA utilizes a standard source follower, which is utilized to drive the large capacitance from the decoder input, as well as to suppress its kick-back noise. The source follower also enables the multiplexing between parallel TIA outputs. AC-coupling is utilized for the inter-stage connections within the TIA, which allows for the photocurrent-independent control of the bias voltage of the DSF. A high-pass frequency corner of 7.9 MHz is achieved, which is sufficiently low for PRBS7 test pattern at the target baud rate.

Extended Data Fig. 4a shows the schematic of the slicer 1 and 3 used in this work, which utilizes a three-stage track-and-regenerate topology [43]. When CK is high, transistors M1−M8 track and pre-amplify the input differential signal. Meanwhile, M11−M14 buffer the outputs of the pre-amplifier and sets the initial voltages of the latch stage (composed of M15−M20). When CK turns low, the latch stage regenerates a rail-to-rail output utilizing the positive feedback from the cross-coupled inverter pair. The latch outputs are buffered by cascaded inverters. In the first half of the clock cycle, the latch stage is disabled (M15 and M16 are turned off), whereas in the second half of the clock cycle, the pre-amplifier stage is disabled as M5−M8 (M9 and M10) are turned off (on), which pulls down the outputs (i.e., $V_{AP}$ and $V_{AN}$) and disables M11 and M12. The major advantage of this slicer topology is the small clock-to-Q delay (i.e. the flip-flop delay) and higher operating speed, due to the tracking function performed in the first half of the clock cycle, which allows the latch stage to start regenerating from a relatively large signal level. Other advantages of this slicer include low power consumption (due to low supply voltage ), higher gain (due to the tracking and pre-amplifying function), and rail-to-rail output levels. In the implemented PAM4 decoder, slicers 1 and 3 sense the upper and lower eye signals (in the PAM4 eye diagram) by comparing the input signal to the threshold voltages (provided to the chip), while the middle slicer (slicer 2, shown in Extended Data Fig. 4b) senses the middle eye signal by detecting the polarity of the input differential voltage. Note that slicers 1 and 3 utilize the double-tail preamplifier design whereas in slicer 2 transistors M3, M4, and M6 are removed and M1 and M2 are used as the input transistors to sense $V_{INP}$ and $V_{INN}$. Such a modification allows for the use of smaller transistors, leading to reduced input capacitance and power consumption.

As shown in Extended Data Fig. 4c, the half-rate MSB and LSB outputs (0° path and 180° path in Fig. 4c) from the PAM4 decoder are converted to 1/8 rate by two 2-to-8 de-serializer modules, which are then used for on-chip BER calculations. The core of the de-serializer consists of three layers of delay flip-flops (DFF) used for two step de-serialization (2-to-4 and 4-to-8) and the output data alignment, respectively. By delay compensating the de-serializer clock path using cascaded inverters, the sampling edges of the first four DFFs are always slightly ahead of data regardless of the data rate. This allows the decoder to have the maximum decision time to avoid metastability and also the de-serializer to have enough set-up and hold times. By utilizing frequency dividers with well-defined output phase relationship [44], the de-serializer avoids any ambiguity in the output data sequence. The inter-stage connections for the clock path and data path in the de-serializer are also delay-compensated to support arbitrary data-rates.

The on-chip BER measurement system utilizes an 8-way parallel PRBS7 generator [41], which is controlled by an off-chip reset signal (RST) such that it first takes the initial pattern from the de-serializer output and then generates input-independent data stream as the target pattern. The RST signal is buffered by multiple cascaded inverters to reduce its rise and fall time, such that all 8 lanes in the PRBS7 generator are initialized within the same clock cycle (in presence of the circuit noise and process variation). The de-serializer outputs are delay-compensated and then compared against the target pattern using an array of XOR gates. Any mismatch, indicating an error bit, is captured by the subsequent on-chip memory module (activated by an off-chip control signal), which sets the measurement duration. Note that by using the on-chip de-serialization and BER measurement system, all electrical I/O signals of the implemented optical receiver (except for the input clock) operate at low frequencies, which significantly simplifies the multi-channel integration and device packaging.

**Acknowledgement**

This work was supported by Defense Advanced Research Projects Agency PIPES program under contract number HR0011-19-2-0016.

**Disclosures**

The authors declare no conflicts of interest.

**Data availability**

Data underlying the results presented in this paper are not publicly available at this time but may be obtained from the authors upon reasonable request.

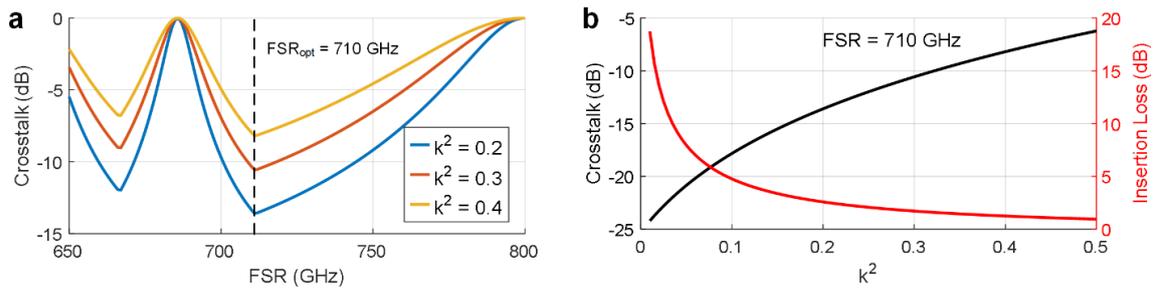

**Extended Data Fig. 1 | Ring resonator loss-crosstalk trade-off simulations. a,** Channel-to-channel crosstalk as a function of the free spectral range (FSR) of a single ring resonator used in one branch of the 8:32 optical demultiplexer (O-DeMux). Four carriers, spaced 1600 GHz apart, are demultiplexed using the drop ports of three cascaded ring resonators coupled to a common waveguide bus, with the final carrier remaining at the through port. **b,** Drop-port insertion loss and crosstalk as a function of the coupling coefficient for a ring resonator with an FSR of 710 GHz.

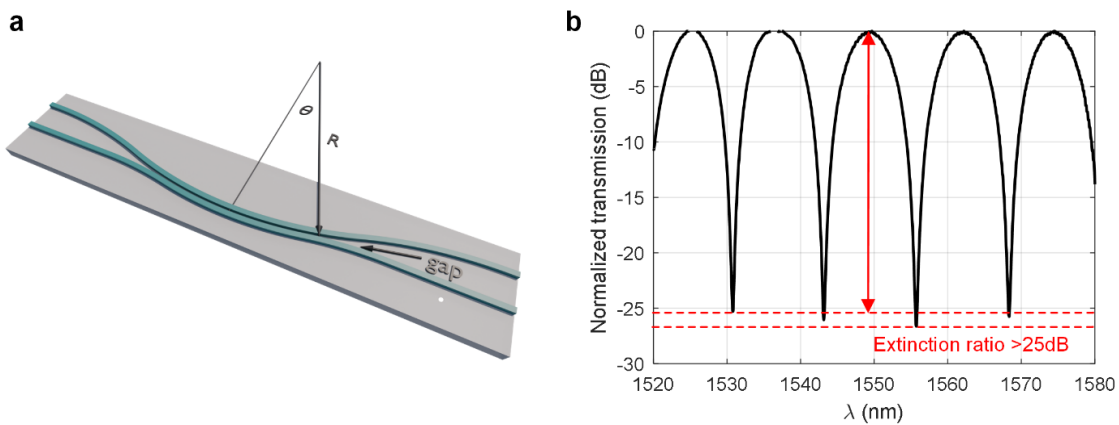

**Extended Data Fig. 2 | Broadband directional coupler. a,** Schematic of the broadband 50% curved coupler used in MZIs. **b,** Measured transmission response of an unbalanced MZI with a 47.5 μm path length difference, achieving a relatively flat extinction ratio over a bandwidth of more than 50 nm.

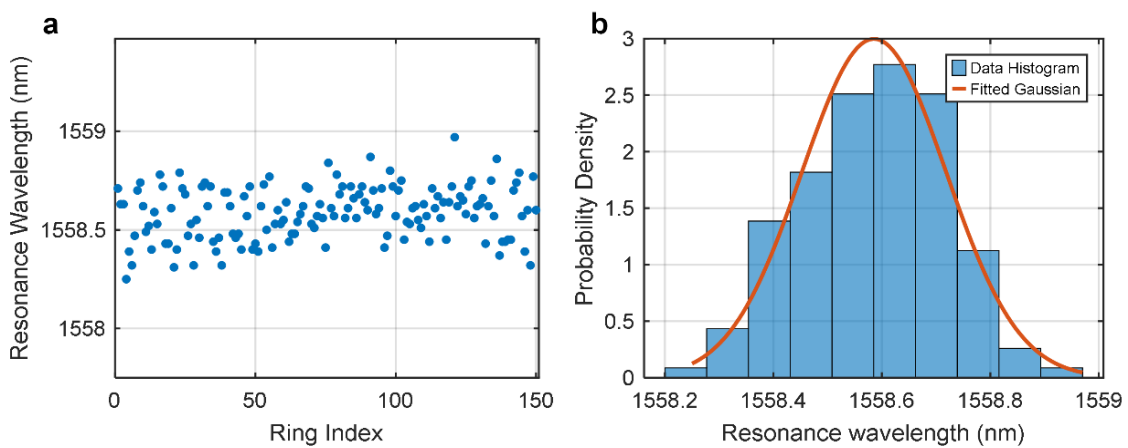

**Extended Data Fig. 3 | Resonance wavelength variation. a,** Scatter plot of the resonance wavelength for 150 capacitively tuned micro-ring resonators fabricated across an area of 1.39 mm × 0.88 mm, showing device-to-device variation. **b,** Probability density function of the resonance wavelengths with a Gaussian fit overlaid, indicating the mean and distribution of the measured resonance wavelengths with a standard deviation of 0.13 nm.

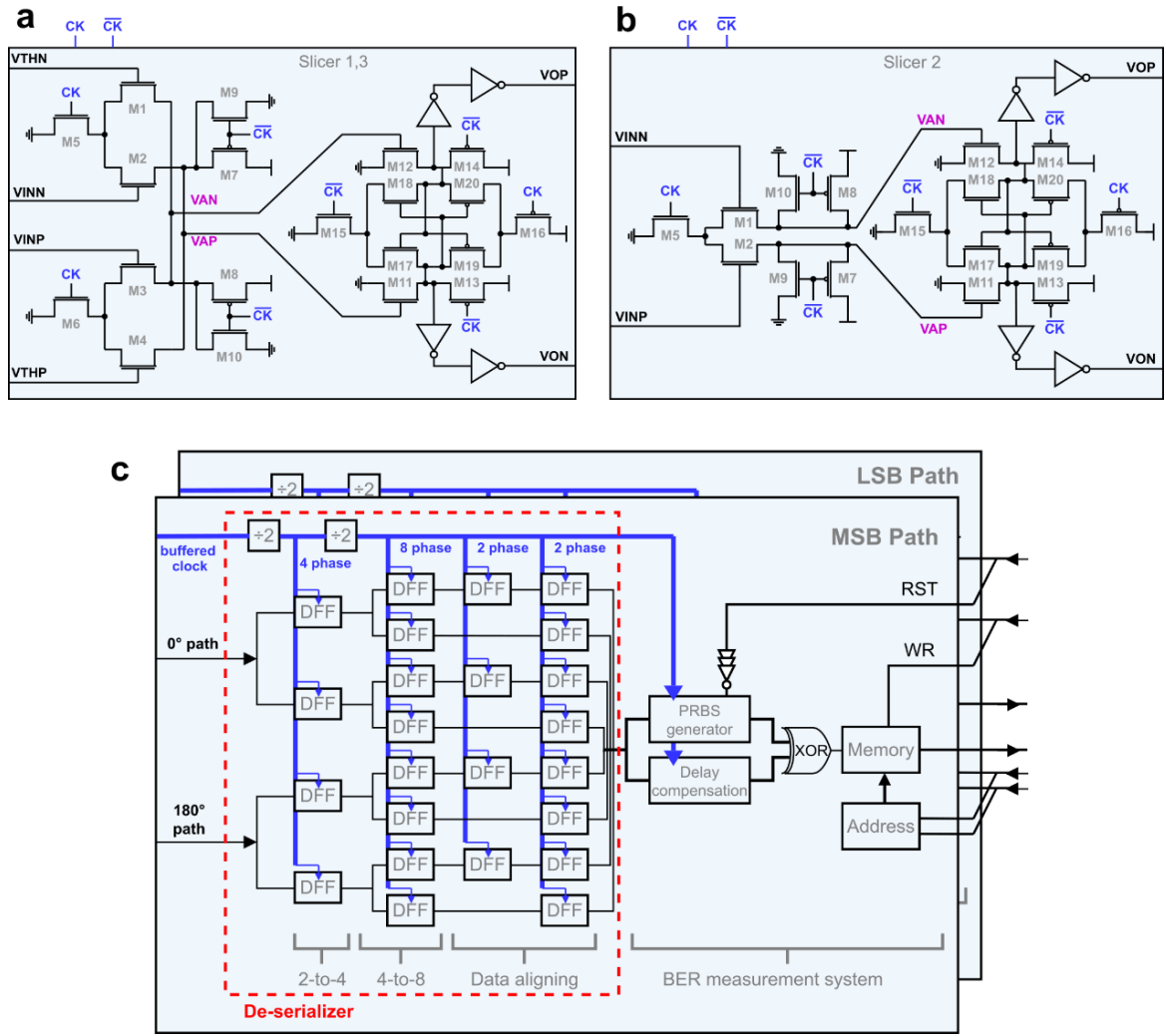

**Extended Data Fig. 4 | PAM4 decoder, de-serializer, and bit-error-rate measurement system. a,** Schematic of the track and regenerate slicers 1 and 3 within the PAM4 decoder in Fig. 4c. **b,** Schematic of the track and regenerate slicer 2 within the PAM4 decoder in Fig. 4c. **c,** Schematic of the de-serializer and bit-error-rate measurement system. DFF: delay flip-flop.

|  | This work | [29] | [30] | [31] | [32] | [33] | [34] | [35] | [36] | [37] | [40] |
|---|---|---|---|---|---|---|---|---|---|---|---|
| Technology | 45nm CMOS | 5nm CMOS | 7nm FinFet | 3nm CMOS | 10nm CMOS | 5nm FinFet | 5nm FinFet | 5nm FinFet | 7nm FinFet | 28nm CMOS | 28nm CMOS |
| Input data | Optical | Electrical | Electrical | Optical | Electrical | Electrical | Electrical | Optical | Electrical | Optical | Optical |
| TIA Included | Yes On-chip | - | - | No (Off-chip) | - | - | - | No (Off-chip) | - | No (off-chip) | Yes (on-chip) |
| WDM | Yes | No | No | No | No | No | No | No | No | No | No |
| Number of channels | 32 | 1 | 1 | 1 | 1 | 1 | 1 | 8 | 4 | 4 | 1 |
| Aggregate data-rate (Gb/s) | 1024 | 224 | 112 | 224 | 112 | 112.5 | 212 | 800 | 425 | 424 | 100 |
| BER | 1e-12 | 1e-6 | 1e-8 | 2.5e-8 | 1e-6 | 7e-6 | 1e-12 | 1e-11 | 2e-5 | 1e-6 | 2.4e-4 |
| Equalization | No | FFE+DFE | FFE+DFE | FFE+DFE | FFE+DFE | FFE+DFE | FFE+DFE | FFE+DFE | FFE+DFE | FFE+DFE | FFE+DFE |
| End-to-end latency | 100ps | N/R | N/R | N/R | N/R | N/R | N/R | N/R | N/R | N/R | N/R |
| Area per channel (mm2) | 0.009 | 0.34 | $0.265^3$ | 0.277 | 0.28 | $0.194^3$ | $0.72^3$ | $0.48^1$ | $0.292^3$ | $0.12^1$ | 0.45 |
| BW density (Gb/s/mm2) | Elec.:3555 All: 2000 | $658^{1,3}$ | $422^3$ | 808.66 | $400^{1,3}$ | $580^3$ | $294^3$ | 208 | $383.5^3$ | 883.3 | 222.2 |
| Power cons. per lane (mW) | 11.77 | $315.2^2$ | $401^2$ | $331^2$ | $448^2$ | 222.2 | $288.3^2$ | - | $338^2$ | 293 | 391 |
| Energy cons. (pJ/bit) | 0.38 | $1.41^2$ | $3.6^2$ | $1.47^2$ | $4^2$ | 1.97 | $1.36^2$ | 4.125 | $3.18^2$ | 2.76 | 3.9 |
| FOM: (BW density) X (energy cons.) | Elec. Only: 9355.2 All: 5263 | $466.6^{2,3}$ | $121.3^{2,3}$ | $550.1^2$ | $100^{2,3}$ | $294.4^3$ | $211.5^{2,3}$ | 50.5 | $120.5^{2,3}$ | 320.0 | 57.0 |

[1] Calculated, [2] Excluding DSP energy consumption, [3] Excluding DSP area

**Extended Data Table. 1 | Comparison with other CMOS PAM4 receivers with aggregate data-rate larger than 100 Gb/s.**